\PassOptionsToPackage{pdfpagelabels=false}{hyperref} 
\documentclass[fleqn,usenatbib]{mnras}
\usepackage{natbib}
\usepackage{caption}
\usepackage{subcaption}
\usepackage{newtxtext,newtxmath}

\usepackage[T1]{fontenc}
\usepackage{ae,aecompl}

\usepackage{graphicx}	
\usepackage{amsmath}	

\title[An FU Ori-type low luminosity protostar]{An outburst and FU Ori-type disk of a former low luminosity protostar}

\author[Mizna et al.]{
Mizna Ashraf,$^{1}$\thanks{E-mail: mizna@students.iisertirupati.ac.in}
Jessy Jose$^{1}$\thanks{E-mail:jessyvjose1@gmail.com},
Ho-Gyu Lee$^{2}$, 
Carlos Contreras Pe\~{n}a$^{3,4}$,
Gregory J Herczeg$^{5,6}$,
\newauthor Hanpu Liu$^{5,6}$,
Doug Johnstone$^{7,8}$
and Jeong-Eun Lee$^{3,9}$
\\
\\
$^{1}$Indian Institute of Science Education and Research (IISER) Tirupati, Rami Reddy Nagar, Karakambadi Road, Mangalam (P.O.), Tirupati 517 507, India \\
$^{2}$Korea Astronomy and Space Science Institute, 776, Daedeok daero, Yuseong-gu, Daejeon, 34055, Republic of Korea\\
$^{3}$ Department of Physics and Astronomy, Seoul National University, 1 Gwanak-ro, Gwanak-gu, Seoul 08826, Republic of Korea\\
$^{4}$Research Institute of Basic Sciences, Seoul National University, Seoul 08826, Republic of Korea \\
$^{5}$ Kavli Institute for Astronomy and Astrophysics, Peking University, Yiheyuan 5, Haidian Qu, 100871 Beijing, People’s Republic of China\\
$^{6}$ Department of Astronomy, Peking University, No.5 Yiheyuan Road, Haidian District, Beijing 100871, People's Republic of China\\
$^{7}$ NRC Herzberg Astronomy and Astrophysics, 5071 West Saanich Road, Victoria, BC, V9E 2E7, Canada\\
$^{8}$Department of Physics and Astronomy, University of Victoria, Victoria, BC, V8P 5C2, Canada\\
$^{9}$SNU Astronomy Research Center, Seoul National University, 1 Gwanak-ro, Gwanak-gu, Seoul 08826, Republic of Korea\\
}

\date{Accepted 2023 December 13. Received 2023 December 12; in original form 2023 September 30}

\pubyear{2023}

\begin{document}
\label{firstpage}
\pagerange{\pageref{firstpage}--\pageref{lastpage}}
\maketitle


\begin{abstract}

Strong accretion outbursts onto protostars are associated with emission dominated by a viscously heated disk, which is characterized by high luminosities. We report the discovery and characterization of a strong mid-IR (3.4, 4.6 $\mu$m) outburst in the embedded protostar SSTgbs J21470601+4739394 (hereafter SSTgbsJ214706). SSTgbsJ214706 has steadily brightened in the mid-infrared by $\sim2$ magnitudes over the past decade, as observed by NEOWISE. Follow-up investigations with the Gemini near-IR spectrograph reveal that SSTgbsJ214706 is a binary system with a spatially extended outflow. The outburst is occurring on the more embedded southeast (SE) component, which dominates the mid- and far-infrared emission from the source. The outbursting component exhibits a spectrum consistent with an FU Ori-type outburst, including the presence of enhanced absorption observed in the molecular bands of CO. The luminosity of the SE component is estimated to be $\sim 0.23\,$ L$_\odot$ before the outburst and $\sim 0.95\,$ L$_\odot$ during the outburst, which is 1 to 2 orders of magnitude fainter than bonafide FU Ori outbursts. We interpret this eruption as an FU Ori-type outburst, although the possibility of brightening following an extinction episode cannot be ruled out. We discuss the implications and potential explanations for such a low-luminosity eruption.

\end{abstract}




\section{Introduction}\label{sec:intro}

The main phase of stellar growth in young stellar objects (YSOs) is expected to occur when the star is deeply embedded in its natal envelope \citep[e.g.][]{1987Lada, 2009Evans,2019Fischer}. During this phase, the rate of accretion is expected to be the dominant contribution to the luminosity of the YSO. The accretion luminosity itself is highly variable at early epochs, as disk instabilities play a key role in stellar mass assembly \citep[see review by][]{2022Fischer}.

Models of stellar growth and pre-main sequence contraction predict that stars should occupy a range of luminosities, however some protostars are underluminous when compared to these expectations. Such {\it low luminosity objects}, characterized by internal luminosities L$_{{\rm{int}}} \lesssim 0.2\,$L$_{\odot }$, represent an extreme case in the observed protostellar luminosity spread. Even without any accretion (an assumption that when time-averaged is not realistic), a protostellar photospheric luminosity of $0.2$ L$_\odot$ corresponds to an upper mass limit of $0.17$ M$_\odot$ at 0.5 Myr in the \citet{2015Baraffe} models, hence many low-luminosity objects may either be proto-brown dwarfs \citep{2013Lee,2016Hsieh,2019Kim} or sub-solar mass protostars that have yet to accrete most of their mass \citep[see, e.g., models by][]{2017Fischer}.

\begin{figure*}
    \centering
    \includegraphics[scale=0.50]{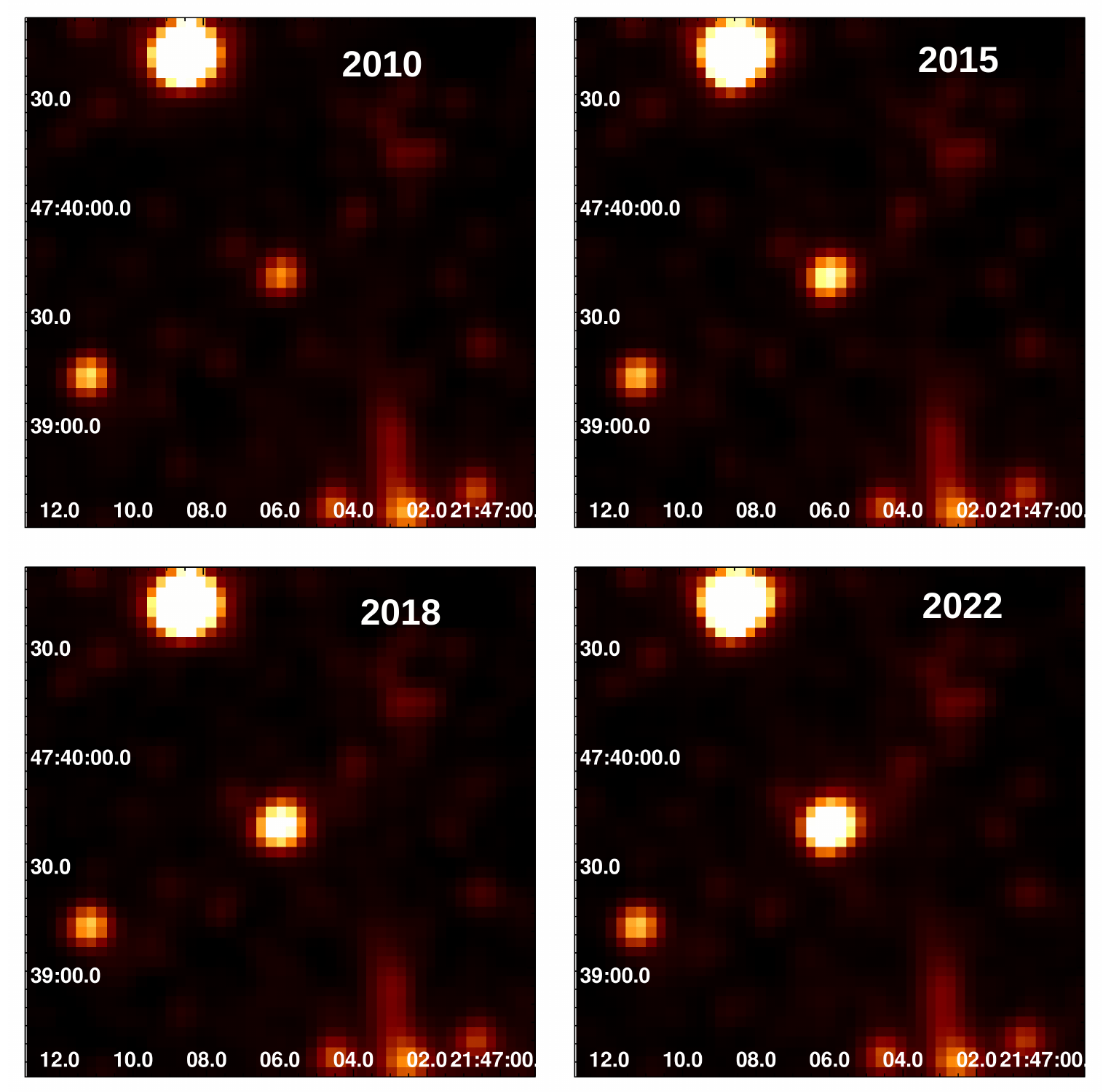}
    \caption{These WISE W2 images taken at various epochs display SSTgbsJ214706, with the source positioned at the center. Each image has been uniformly scaled for easy comparison, revealing a clear brightening event over 12 years.}
    \label{NEOWISE}
\end{figure*} 


A plausible explanation for a lack of accretion luminosity is that accretion is highly episodic. Modeling of chemical abundances in protostellar envelopes as well as the luminosity functions of protostars predicts that bursts are a common mode of accretion in protostars \citep{2014Bae,2015Visser,2019Hsieh}. During most epochs of quiescence, the accretion luminosity would be less than the stellar luminosity, but during short periods of strong stellar growth, accretion could bring substantial near-IR luminosity excess \citep{2022Liu}. Indeed, a significant fraction of low luminosity protostars exhibit outflow signatures believed to result from prior accretion bursts \citep{2020Anderl, 2018Hsieh, 2016Kim, 2023Dutta}. Direct evidence for such episodes are found in FU Ori-type outbursts, which erupt from  quiescent accretion rates of $\sim {10}^{-8} - {10}^{-7} \,{M}_{\odot }\,{\mathrm{yr}}^{-1}$ to reach as much as ${10}^{-4} \,{M}_{\odot }\,{\mathrm{yr}}^{-1}$ \citep{1996Hartmann}. However, discoveries of FU Ori-type outbursts are rare; their discovery has historically been possible through optical surveys, which miss the critical protostellar phase when the main stellar mass assembly occurs.

Recent observations of YSO variability in the infrared and sub-millimeter domains have revealed an increased frequency of outbursts among stars in their early stages of evolution \citep{2019Fischer, 2020Guo, 2021Lee, 2021Park, 2022Zakri, 2023Carlos}. 
The \citet{2021Park} survey of NEOWISE variability in nearby star-forming regions identified  SSTgbsJ214706, a source in IC 5146 (cluster distance of 
783$\pm$36 parsec, as measured by \citealt{2018Zucker} from Gaia DR2 \citealt{2018Gaiadr2}) previously classified as a low-luminosity protostar \citep{2021Kim_M}, as a candidate eruptive variable with a long-term rise of $\sim 2$ magnitudes over 12 years (see Figure \ref{NEOWISE}). Prior to the outburst, this object was classified as a Class I YSO based on its Spectral Energy Distribution (SED) \citep{2010Heiderman}. 

In this paper, we present the observations of SSTgbsJ214706, with imaging that separates the object into two distinct components, and Gemini/GNIRS near-IR spectroscopy supplemented with historical photometry to characterize both components and the burst. The SE (South-East) component of SSTgbsJ214706, which is more deeply embedded, is responsible for the ongoing eruption and has a spectrum that is characteristic of FU Ori-type objects, whereas the NW (North-West) component is a G star and does not appear to be strongly accreting. SSTgbsJ214706 has its outburst luminosity an order of magnitude lower than known FU Ori-type objects, making it unique.

The paper is organized as follows: Section 2 describes the observational details, while Section 3 reviews previous studies on SSTgbsJ214706 and examines the possibility of binarity. Section 4 presents the photometric analysis, including an overview of the observed light curve and color variations as well as analysis of the SED. Section 5 includes the spectral analysis, and Section 6 presents the discussion.

\section{Observations}

\begin{figure*}
    \centering
    \includegraphics[scale=0.59]{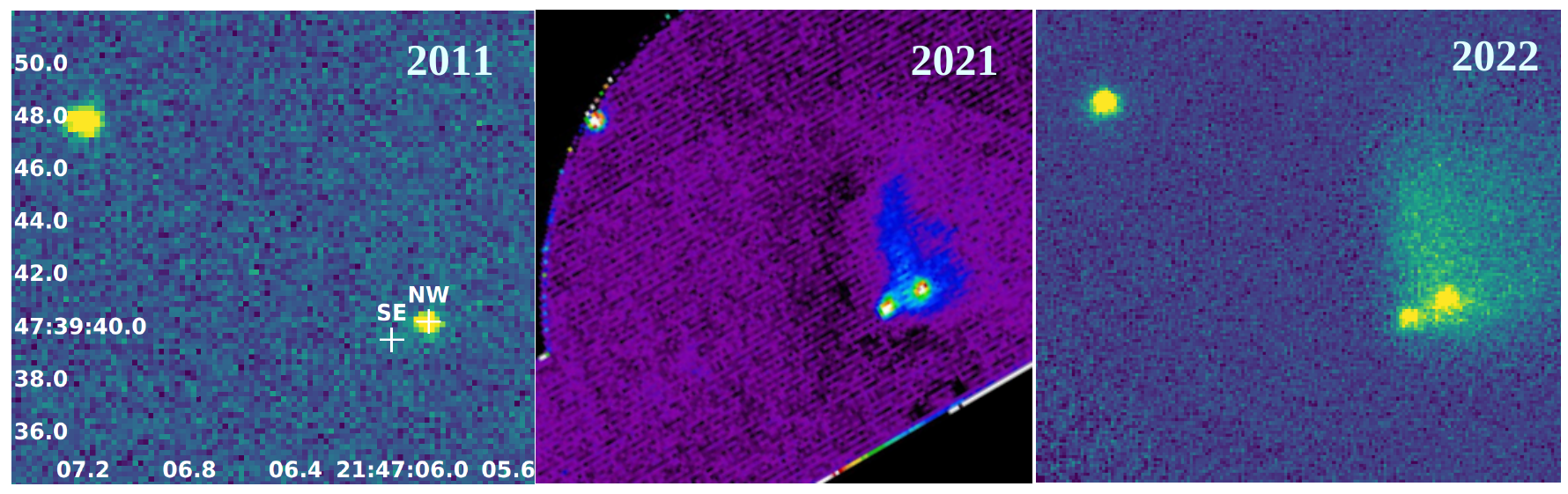}
    \caption{Images of SSTgbsJ214706 taken at three different epochs: UKIDSS K band image from August 2011 (left panel), GEMINI/GNIRS acquisition image from August 2021 (middle panel), and IRTF-spex acquisition image from July 2022 (right panel). The components NW and SE (see text for details) are marked in the left panel. The 2011 image does not show the presence of the second component, SE, which becomes apparent in the more recent GEMINI and IRTF images. Recent images from IRTF and Gemini also reveal an increased presence of extended emission in the star's vicinity.} 
    \label{aquisition image}
\end{figure*}

We carried out spectroscopic observations in the wavelength range 1.45 - 2.45 $\mu$m on 9 August 2021, with GNIRS on the 8.1 m Gemini North telescope located at Mauna Kea, Hawaii (programme GN-2021A-Q-106, PI Johnstone). We used the 31.7 l/mm grating (long-slit mode), a 0.3\arcsec slit, and the cross-disperser with the short camera to achieve a resolution of R=$1700$ at 2.2 $\mu$m. The observations consist of 16$\times300$~s exposures with several series of ABBA nodding at different positions on the slit to eliminate the sky background. Bright A0V standards were observed for telluric calibration. 

The acquisition image taken using GNIRS (see Figure \ref{aquisition image}) indicates that the target consists of two point sources, with diffuse nebular emission extending $\sim 5 \arcsec$ from between the two stars to the north and west. This diffuse emission is visually more associated with the target towards west in the image. However, measuring the flux and positions of the two components from the Gemini acquisition image was not feasible as there is only one background source available for WCS correction and flux calibration.

During observations, the nodding length along the slit was less than the size of the extended emission. The reduction steps with standard procedures using the GEMINI package in IRAF resulted in spectra contaminated with diffuse emission. Hence we carried out a manual extraction with python considering only A-B pairs to find the peaks of two stars and vertically extract 5 pixels centered at the peaks. The contribution from extended diffuse emission was removed from the spectrum of the source by choosing a sky region about 1 to 4 times of the separation distance between two stars. 

We also obtained K band acquisition images from the 3.2 m IRTF in Hawaii on 22 July 2022 using the Spex instrument. Since the target is quite faint, we were unable to acquire the spectrum. The acquisition frames were taken in the K-band, which has a field of view of 1\arcmin\ and a plate scale of 0.12\arcsec\ per pixel. The seeing during  the night was $\sim$0.62\arcsec\ and hence both the above components were resolved. Five frames of 20 seconds each were combined. This image was WCS converted using five reasonably bright common background sources from UKIDSS K band archival images \citep{2008Lucas}. The UKIDSS image also serves as an indicator for the identification of the source component under outburst  (see Figure \ref{aquisition image}). 

In addition, we use several archival photometric data, including Spitzer-IRAC/MIPS photometry \citep{2015Dunham}, WISE and NEOWISE time series data from 2010 to 2022 including 21 epochs \citep{2014Cutri}, Herschel PACS photometry \citep{2016Kim}, JCMT SCUBA-2 450 AND 850 $\mu$m \citep{2017Johnstone} data and the sixth data release of the UKIDSS Galactic plane survey \citep[GPS,][]{2008Lucas}.

\begin{figure*}
    \includegraphics[scale=0.35]{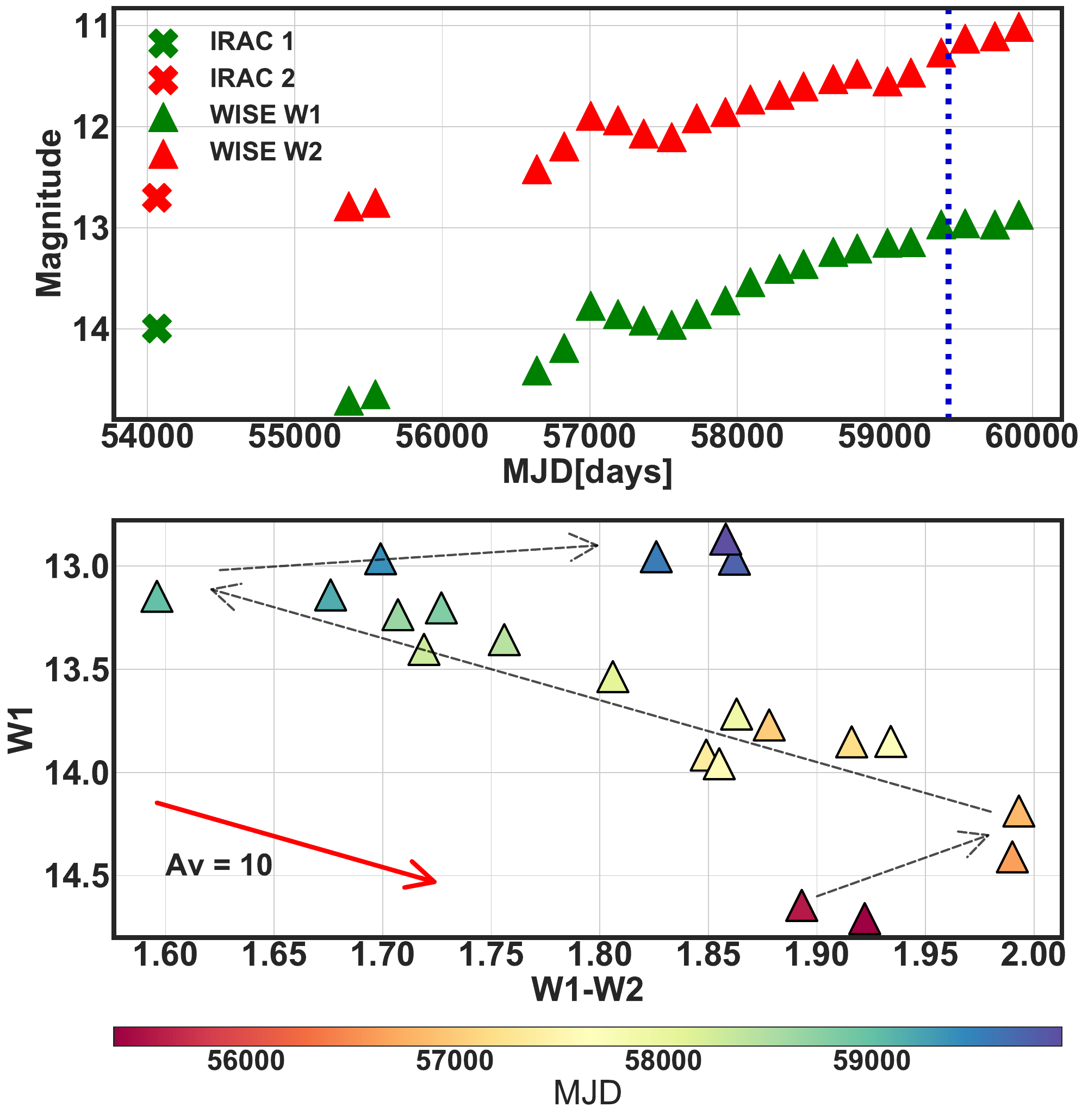}
    \caption{ Top: The light curve for SSTgbsJ214706 (combined light curve for both components), covering a 16-year period from 2006 to 2022 and constructed using one epoch of IRAC 3.6 and 4.5 $\mu$m magnitudes (2006), two epochs of the AllWISE W1 and W2 magnitudes (2010), and 19 epochs of the NEOWISE W1 and W2 magnitudes (from 2013 to 2022). The green and red symbols indicate brightness at $ \sim 3.4$ and $\sim 4.6\, \mu$m respectively. The epoch at which the Gemini spectrum was obtained is indicated by the blue dotted line. The light curve reveals a gradual increase in brightness since 2010, spanning a duration of 13 years. Bottom: W1 versus W1-W2 colour-magnitude diagram constructed using WISE and NEOWISE data from 2010 to 2022. The red arrow mark shows the reddening vector representing a reddening of A$_{V}=10$~mag, using the extinction law of \citet{2019Wang}. The black dashed line indicates the trajectory of the object's colour evolution.}
    \label{lc}
\end{figure*}

\section{Source Overview and Binarity} \label{Source}

SSTgbsJ214706 ($\alpha$ = 21:47:06.01, $\delta$ = +47:39:39.4), is a young stellar object that was first identified by \citet{2008Harvey} in the Spitzer Space Telescope Gould Belt Survey, associated with the cluster IC 5146. The properties of SSTgbsJ214706 have been previously measured under the assumption that it was a single star. The mid-IR spectral index of $\alpha$ = 0.40 leads to the classification of SSTgbsJ214706 as a Class I YSO \citep{2010Heiderman,2015Dunham}. The bolometric luminosity was estimated from Spitzer IRAC/MIPS photometry as 0.06 L$_{\odot}$, after correcting for the updated distance of 783 pc \citep{2015Dunham}. \citet{2021Kim_M} classified the YSO as a very low luminosity object (VeLLO) based on Spitzer and Herschel photometry with a luminosity of $\sim 0.23\,$L$_\odot$ (revised for the updated $783$ pc distance by \citealt{2015Dunham}). \citet{2017Johnstone} confirmed the presence of an envelope by associating the infrared source with a  submillimeter clump (at both 450 and 850 $\mu$m) using JCMT Gould Belt Survey data.

The K-band acquisition images obtained from GNIRS and IRTF Spex (Figure \ref{aquisition image}) reveal that the source is comprised of two distinct components, which we call SE and NW to denote their relative positions. We assume that both components form a binary system within IC 5146, given the common occurrence of binarity in embedded protostellar systems \citep[e.g.][]{2008Connelley,2022Tobin}. However, since the object is not detected in Gaia, we lack sufficient data to definitively verify binarity and rule out the possibility that one component is a background object. The spatial coordinates for SSTgbsJ214706 SE and NW components are determined as (21:47:05.92, +47:39:40.1) and (21:47:06.06, +47:39:39.4), respectively. These coordinates were obtained by fitting a 2D Gaussian profile to the WCS converted IRTF K-band acquisition image, using the IMEXAM task in IRAF. This corresponds to a projected separation between the two components of 1.58\arcsec (or $\sim 1250$ AU), hence it is plausible to infer that these components are part of a binary system. 

Figure~\ref{aquisition image} shows that only one of the components was visible during the UKIDSS GPS K-band image obtained in 2011. The centroid of this component is consistent with the location of SSTgbsJ214706 NW, as measured from the recent IRTF image, and hence we conclude that SSTgbsJ214706 NW was the source detected with UKIDSS. The detected source NW has a brightness of 16.2 mag in K band from UKIDSS DR6 archival photometry \citep{2008Lucas}.

\begin{figure}
    \centering
    \includegraphics[scale=0.50]{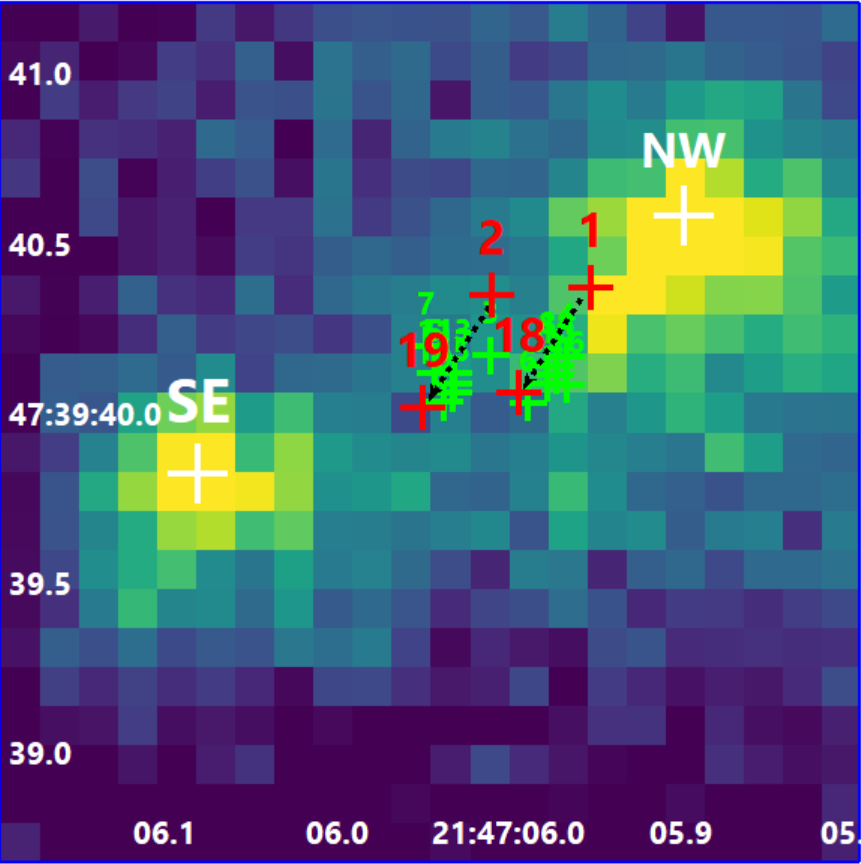}
    \caption{The center of brightness of the system, as observed in the WISE W2 images from 2010 to 2022. The background image is the the IRTF acquisition image during outburst phase (2022). The white crosses represent the coordinates of NW and SE derived from the IRTF acquisition image. The red crosses represent the center of brightness measured from the initial epochs (2010, marked as epoch number 1 and 2) and the most recent epochs (2022, marked as 18 and 19) of WISE W2 images. The green crosses represent intermediate epochs with corresponding epoch numbers labelled. A noticeable trend is observed as the system brightens, with the center of brightness gradually shifting towards SE.}
    \label{shift}
\end{figure}

\section{Photometry of the Outburst}\label{sec:photo}

SSTgbsJ214706 is considered as a binary system with a radial separation of 1.58\arcsec, with the SE component undergoing an outburst. Hence we separate as best as possible the photometry for individual components and concentrate our analysis on the properties of the outbursting source.

The available photometric data for SSTgbsJ214706 includes Spitzer-IRAC/MIPS photometry \citep{2008Dunham}, WISE and NEOWISE time series data from 2010 to 2022, encompassing 21 epochs \citep{2014Cutri}, Herschel PACS photometry \citep{2016Kim}, JCMT SCUBA-2 450 and 850 $\mu$m imaging \citep{2017Johnstone}, and the sixth data release of the UKIDSS Galactic plane survey \citep{2008Lucas}. This photometry represents the total system brightness, summing the contributions from both components. We use the photometric observations acquired on or before 2010 to build the SED and extract the pre-outburst properties of the source. For the photometric analysis during the outburst phase, we use 19 epochs of NEOWISE mid-infrared photometry (W1 and W2) spanning from 2013 to 2022 and K-band photometry derived from the IRTF acquisition image in 2022. 

\subsection{The Outburst Discovered in the Mid-Infrared}

 The mid-IR emission from SSTgbsJ214706 shows a gradual brightening over the past 12 years of 1.78 mag at 3.6 and 1.84 mag at 4.5 $\mu$m, as revealed by NEOWISE time series observations (see the lightcurve given in Figure \ref{lc}). The e-folding rise time measured with respect to the beginning of rising phase (2010 epoch), is 2500 days ($\sim$7 years). The $3.6$ and $4.5$\, $\mu$m Spitzer/IRAC brightness during 2005 was 13.99 and 12.69, which is slightly higher than the WISE brightness of 14.70 and 12.78 during 2010, indicating that SSTgbsJ214706 faded slightly between these two epochs.

The mid-Infrared images from Spitzer and WISE reveal the existence of a slightly extended source along the same axis as the binary, although the spatial separation is smaller than the angular resolution of Spitzer and WISE. We calculated the centroid of the source in the NEOWISE images for each epoch to detect any potential trends. To account for slight offsets in right ascension in alternate epochs, the NEOWISE epochs were divided into two sets. The centroid was determined by measuring the peak of a 2D Gaussian profile fit to the image using the IMEXAM task in IRAF for each epoch \citep[see also][]{2022Yoon}. Based on the analysis of NEOWISE W2 images, the centre of brightness shifts by 0.4\arcsec towards the SE component as the brightness increases (see Figure \ref{shift}), indicating that SSTgbsJ214706 SE is the source experiencing the outburst. Although $\sim 0.4\arcsec$ is much smaller than the pixel scale of WISE, the shift is systematic in both right ascension and declination and is consistent in both sets of epochs (see Figure \ref{sys} in Appendix). However, W1 images do not show any systematic trend, likely due to a low signal-to-noise ratio (SNR).

Figure \ref{lc} illustrates the W1 versus W1-W2 color-magnitude diagram. The observed variation in colours of the source is not unidirectional. The source becomes redder during the initial phase, likely caused by the increased contribution from the redder source SE. The source then becomes bluer with increased brightness, consistent with increased viscous heating in the inner disk (i.e., from a clump of material passing through it) or perhaps an increase in magnetospheric accretion, \citep[e.g.][]{2022Liu,2023Cleaver}, and it may also be influenced by a decrease in extinction. The source again becomes redder as the W1 magnitude increase slows near $\sim13$, possibly due to the heating of cooler regions as the outburst propagates outwards \citep{1994Bell}.

\subsection{K-band brightening of the outbursting source}

The K-band magnitudes for SSTgbsJ214706 NW and SE are determined to be 15.94 and 16.11, respectively, based on PSF photometry on the IRTF/SpeX K-band acquisition images (July 2022) using IRAF. The UKIDSS K-band magnitude of source NW was recorded as 16.21 prior to the outburst \citep {2008Harvey}. The observed increase in brightness of $\sim 0.27$ mag may be attributed to the the extended nebulosity, which appears to have manifested recently, and potential differences in the filter transmission curves between IRTF and UKIDSS. 

The 5$\sigma$ sensitivity limit of the UKIDSS GPS survey is given as $K=18.1$~mag; however, this limit can vary depending on the region covered by the survey \citep{2008Lucas}. To estimate the sensitivity limit in the region where our source is located, we examined the UKIDSS K-band photometry of objects around SSTgbsJ214706. We queried the Wide Field Camera (WFCAM) Science Archive \citep[WSA,][]{2008Hambly} to select stars (objects with image profile classifier {\it mergedClass}$=-$1 ) within an arbitrary radius of 10\arcmin. From the magnitude distribution of these sources, we estimate a 5$\sigma$ limit of K$=18$ mag for this region. This implies a minimum threshold for the increase in brightness of SE, estimated to be $\Delta K\simeq1.9$ mag.

\subsection{Individual Photometry and SED}\label{sec:sed}

The pre-outburst SED of SSTgbsJ214706 SE is constructed using the photometric measurements obtained on or before 2010. This includes the Spitzer-IRAC/MIPS photometry from 2005, WISE All-Sky data (2010), Herschel PACS photometry (2010), JCMT SCUBA-2 450 and 850 imaging (2010) and the sixth data release of the UKIDSS GPS (2005). All of these measurements correspond to the total flux of the system as the object had not previously been identified as a binary. Hence, to determine the pre-outburst photometry of the individual components, we adopted the following approach. 

\begin{figure}
    \centering
    \includegraphics[scale=0.6]{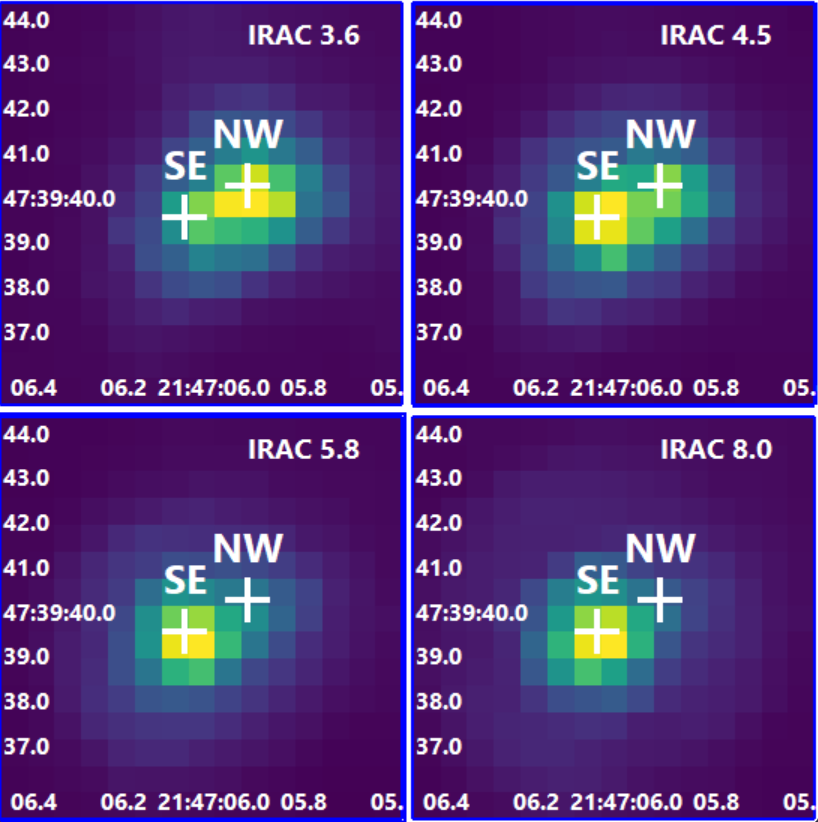}
    \caption{Spizer-IRAC multi-band images from pre-outburst epoch in 2006. The white crosses represent the spatial coordinates of the two components NW and SE derived from the IRTF-Spex K band acquisition image. The images clearly indicate SE component becoming brighter at longer wavelengths.}
    \label{irac}
\end{figure}

As shown in Figure \ref{irac}, while both the SE and NW components contribute to the image flux in Spitzer 3.6, 4.5, and 5.8 $\mu$m bands, the SE component gradually outshines with the increase of wavelength and appears to dominate at $\lambda=8\,\mu$m.  Therefore, we evaluate the SE and NW photometry individually for $\lambda \leq 5.8\,\mu$m and assume that photometry in redder bands ($\lambda \geq 8\,\mu$m)  are solely due to the SE component. The Spitzer IRAC 3.6, 4.5 and 5.8 $\mu$m images (Figure \ref{irac}) show an extended source.  Individual photometry for the SE and NW components are estimated from these images using the PSF photometry method in DAOPHOT by fixing their wcs positions (obtained from IRTF aquisition image). A complete list of the individual photometric measurements for both SE and NW components is provided in the appendix (Table \ref{table:1}).

The SEDs (Figure \ref{sed}) indicate that the SE component is deeply embedded in an envelope, seen through very high extinction.  The NW component is brighter than the SE component in the near-IR, indicating that it is less embedded and has lower extinction.  The NW component does not clearly have excess IR emission from a disk, although an excess also cannot be ruled out.

We employed the SED fitting algorithm in \citet{2006Robitaille} using the updated model grid \citep{2017Robitaille} of young stellar objects to analyze the pre-outburst properties of the outbursting component SE. To determine the most appropriate model for our source, we followed the selection criteria outlined by \citet{2017Robitaille} based on the condition ${\chi}^2 - {\chi}_{\rm best}^2 \le 3 n_{\rm data}$ using a range of distances from 750 to 800 parsecs \citep{2018Zucker} and the extinction curve from \citet{2010Forbrich}. The interstellar extinction range was set between 0 and 100 mag, taking into account both intrinsic dust from the circumstellar environment and extinction from the interstellar medium. The SED fit yielded a best fit model comprising a central star with a temperature of $\sim4000$ K, an envelope, a disk component, bipolar cavities, an ambient medium, and an inner hole (see Table \ref{sedparams} in Appendix). The best fit extinction value is $ A_v \sim 65 $ mag. The pre-outburst bolometric luminosity of the source is measured to be $0.25\,$L$_\odot$ by integrating the flux of the best-fit model across the entire wavelength range. It is important to note that the models by \citet{2017Robitaille} include a passive disk component but do not explicitly consider accretion, potentially introducing inaccuracies in the estimation of luminosity and other parameters.

\begin{figure}
    \centering
    \includegraphics[scale=0.31]{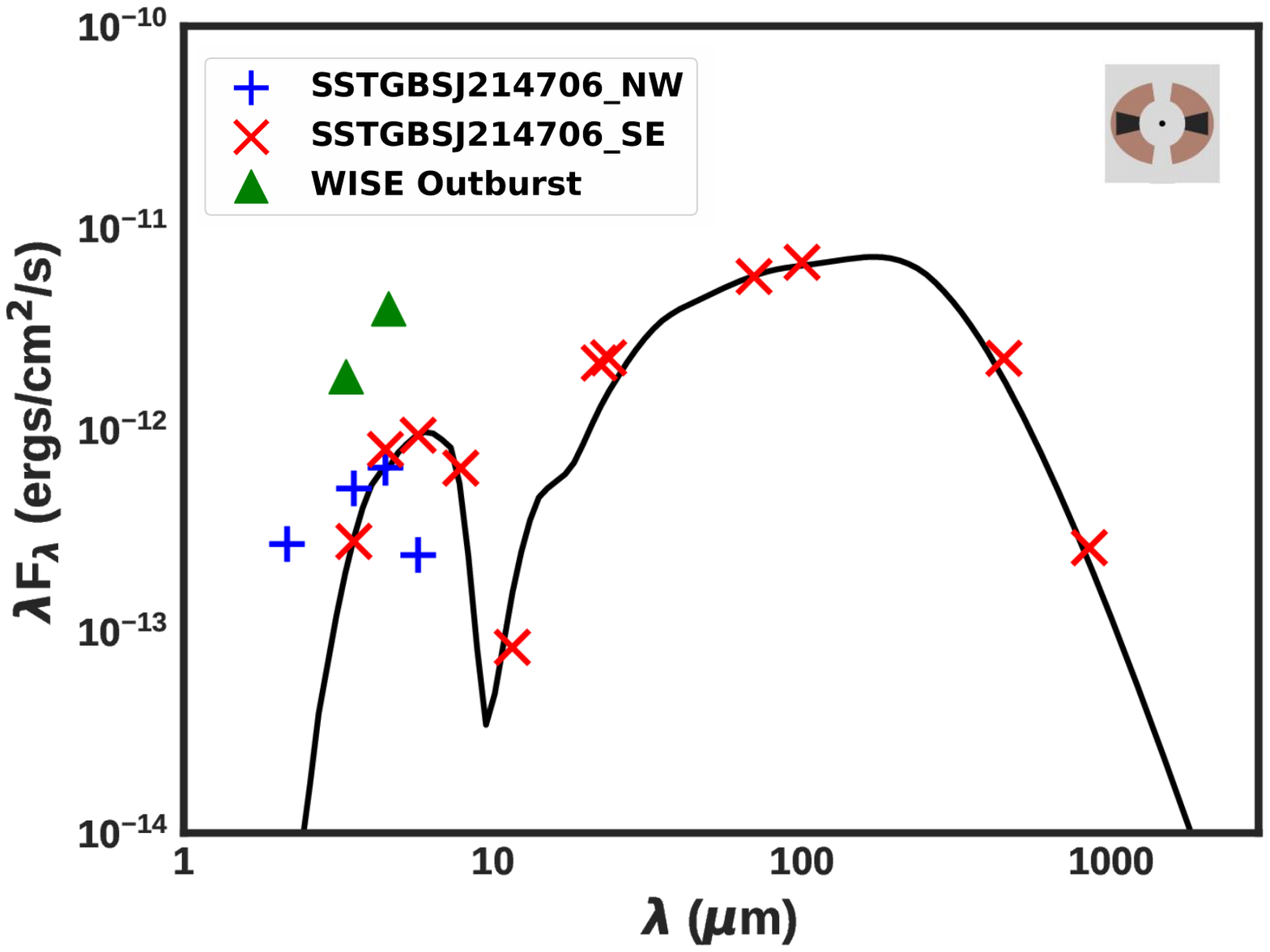}
    \caption{SED fit with the \citet{2017Robitaille} YSO models. The best-fit stellar model is given in the upper right hand side. The red crosses representing the individual fluxes for the SE component obtained prior to 2010 are used for the SED fitting. The solid black line shows the best-fit YSO model. Green triangles represent the latest epoch WISE photometry (2022). Blue crosses are the individual photometry obtained for NW from UKIDSS and IRAC images (2005).}
    \label{sed}
\end{figure}

\begin{figure*}
     \centering
     \includegraphics[scale=0.50]{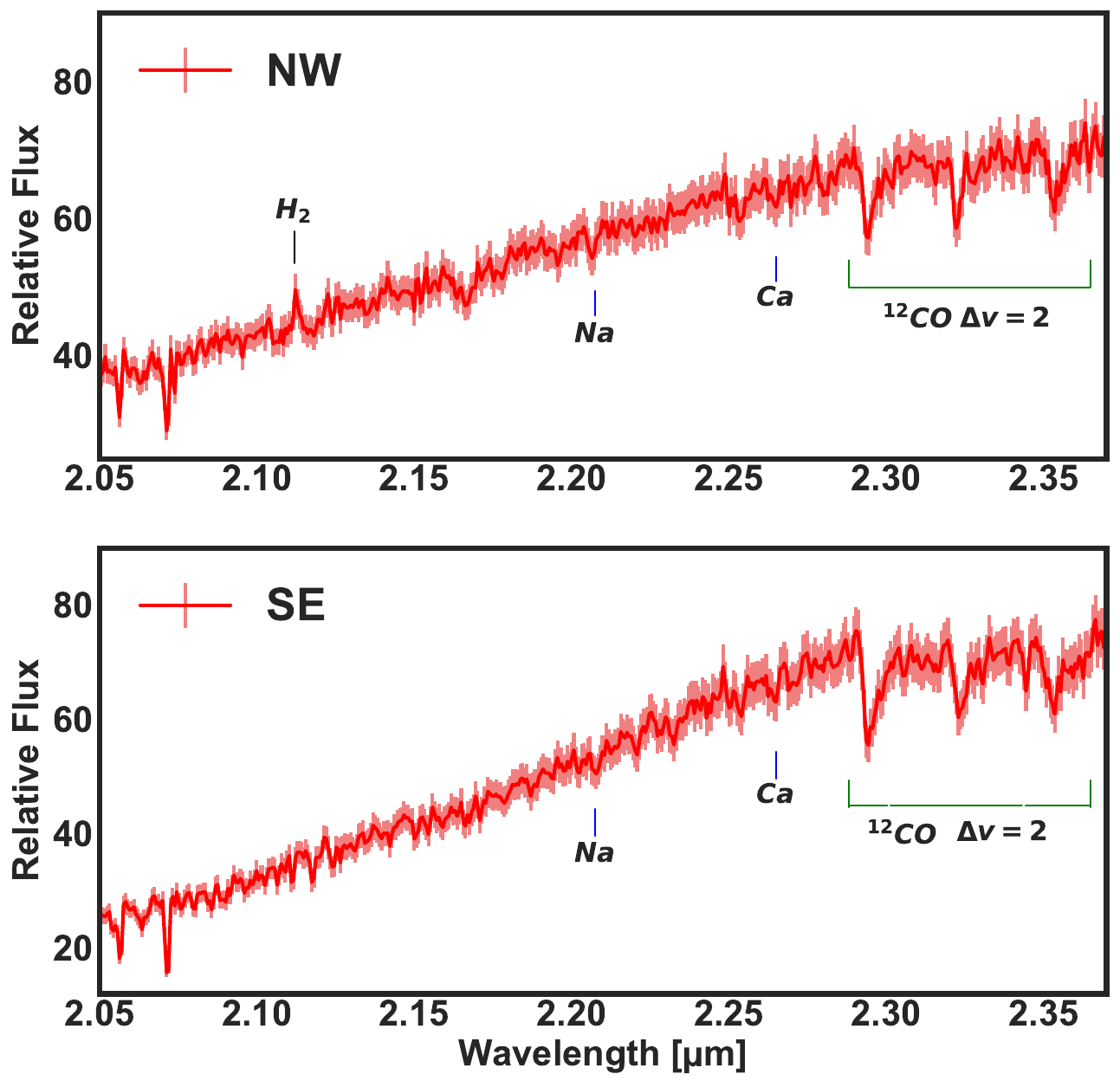}
\caption{ K band spectrum of NW and SE obtained using Gemini/GNIRS. The prominent emission/absorption features present in the spectrum are marked and labelled.}
\label{spectrum}
\end{figure*}

\begin{figure*}
     \centering
    \includegraphics[scale=0.32]{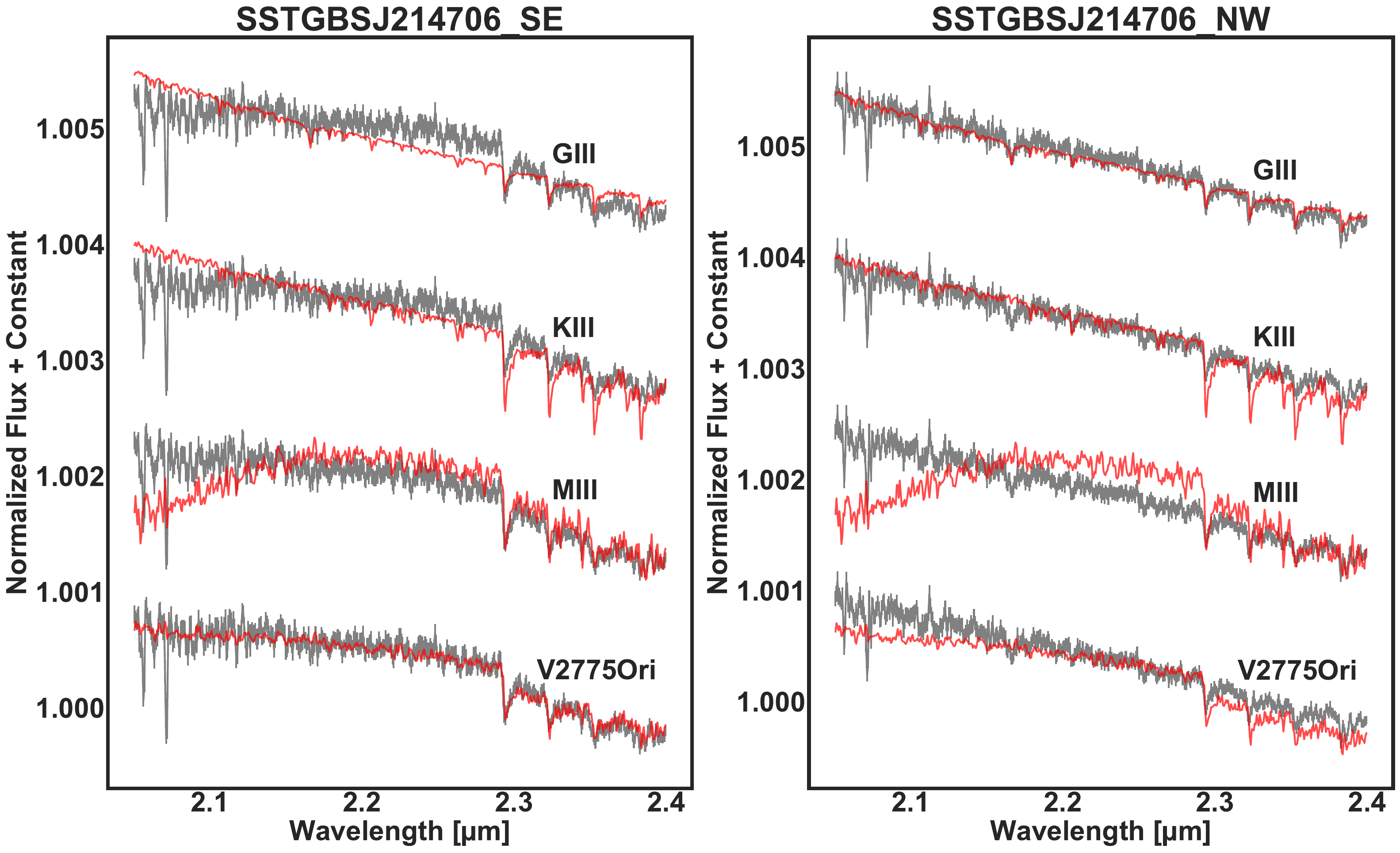}
\caption{K band spectrum of NW and SE (grey) along with the best fit templates (red). The spectra are corrected for an extinction of $A_V=50$ mag. The spectral type of SSTgbsJ214706 SE matches M9III at the longer wavelength end due to its enhanced absorption in CO, which are typical of FU Ori-type objects, whereas the spectrum of NW best matches the G7III spectral type. The SE spectrum shows close resemblance to the spectrum of bonafide FU Ori-type object V2775 Ori.}
\label{sp}
\end{figure*}

\begin{figure}
    \centering  
    \includegraphics[scale=0.36]{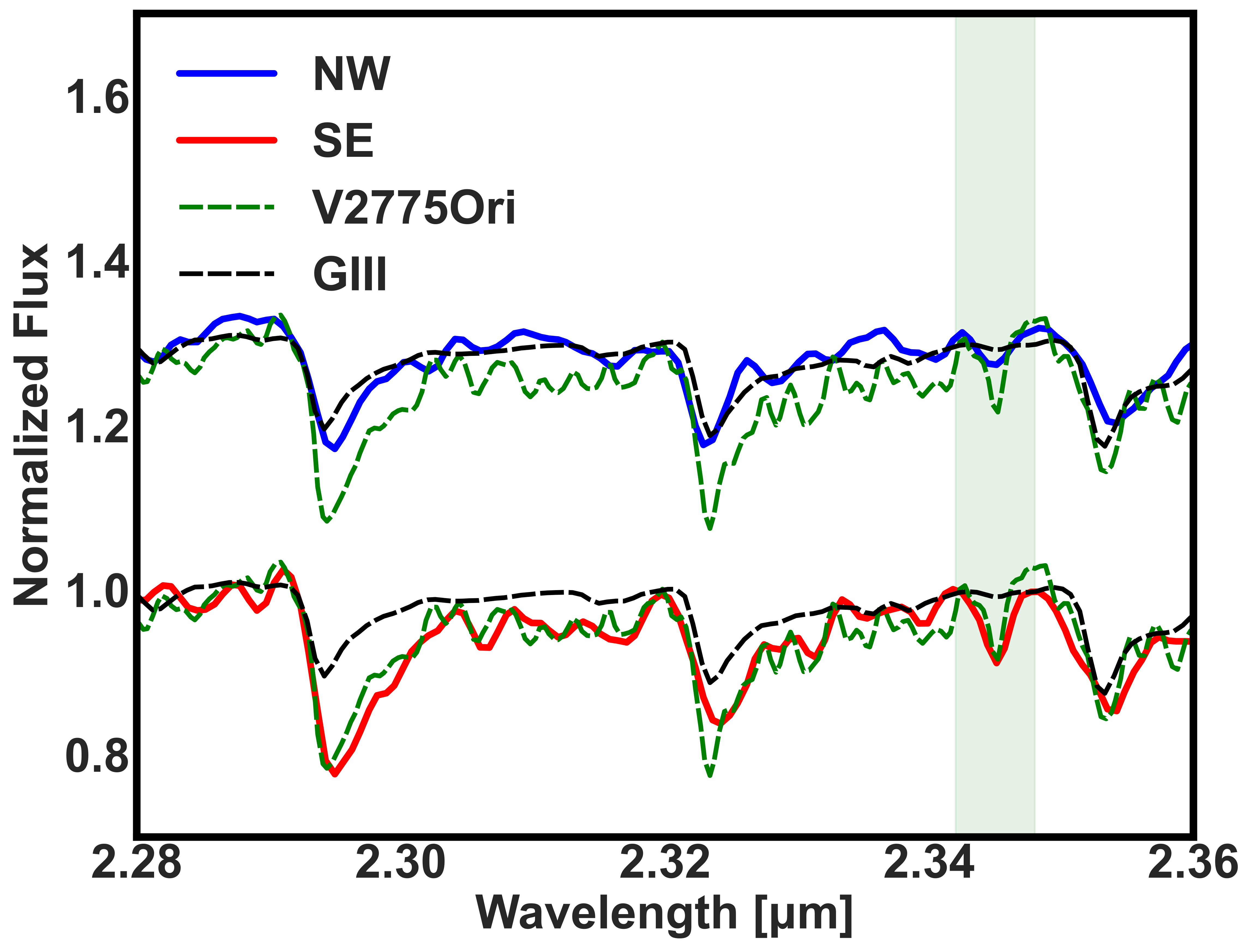}
    \caption{The CO 2-0 absorption feature ( $\sim  2.29 \mu$m) of SE closely resembles that of the bonafide FU Ori-type object V2775Ori, whereas the spectrum of NW matches with a G type star. The green shaded region highlights the absorption feature observed in the SE component at $\sim 2.345\,\mu$m similar to that in V2775\,Ori , which is weaker in NW.}
    \label{co}
\end{figure}

\subsection{Pre-outburst and outburst luminosities}

The pre-outburst luminosity of SE is estimated using the correlation between the 70 $\mu$m flux and the internal luminosity of a protostar, given by \citet{2008Dunham}. 
\begin{equation}
  \hspace{2 cm}  L_{int} = 3.3 \times 10^8 \times F_{70}^{0.94} \times L_\odot.
\end{equation}

Assuming that the 70 $\mu$m flux of 134 mJy, obtained from Herchel/PACS observations in May 2010 \citep[Obs ID:1342197305]{2016Kim} prior to the outburst, solely originates from the SE component of SSTgbsJ214706, we calculate a total pre-outburst luminosity of $0.23\, $L$_{\odot}$. This estimate aligns with the luminosity measurements reported by \citet{2021Kim_M} as well as the luminosity of $0.25\,$L$_\odot$  from the best-fit SED model of SE obtained from \citet{2017Robitaille}. 

\citet{2020Contreras} found that during an outburst, the increase of the mid-IR flux and luminosity of deeply embedded protostars follow a relationship F$_{IR} \propto$ L$^{3/2}$. We estimate the outburst luminosity of SSTgbsJ214706 SE on the assumption that the brightness of NW component remains constant.

The brightness measurements for the SE and NW components obtained from Spitzer/IRAC 4.5 images during the November 2006 epoch, are 0.97 mJy and 1.20 mJy, respectively. This implies that the SE component contributes $\sim$ 55\% of the total flux at 4.5 $\mu$m in the pre-outburst phase.
Using the pre-outburst WISE W2 flux from the initial epoch (June 2010), the contribution from SE component would be 0.73 mJy (55\% of 1.32 mJy). Assuming the NW component has a constant flux of 0.59 mJy (45\% of 1.32 mJy) in WISE W2, the W2 flux of the SE component during the outburst (latest epoch November 2022), can be calculated as 6.17 mJy ( Combined; 6.76 mJy - NW; 0.59 mJy). Thus the ratio of outburst to pre-outburst fluxes for SE component is calculated to be $\sim$ 8.45. The outburst luminosity can be estimated as $8.45^{(2/3)} \times 0.23$ L$_\odot$, which is $\sim$ 0.95 L$_\odot$.

\section{Characterizing the outburst with infrared spectroscopy}

Protostars with high accretion rates, including FU Ori-type outbursts, exhibit optically thick disk photospheres heated internally by viscous dissipation \citep[e.g.][]{2007Zhu}. The radiation caused by viscous heating of the disk midplane is subsequently absorbed by the relatively cool atmosphere of the disk, thereby resulting in absorption in the CO overtone bands and water vapour, while the Na and Ca absorption features are weaker than expected for a stellar photosphere \citep[e.g.][]{2018Connelley}. FU Ori-type objects typically do not show emission lines, with Br-$\gamma$ either missing entirely or extremely weak. The near-infrared spectral characteristics are best matched with K-M supergiant atmospheres, with effective temperatures of 2000-3000\,K (see, e.g., review by \citealt{1996Hartmann}).

Figure \ref{spectrum} shows the K-band spectra of SSTgbsJ214706 NW and SE. The spectrum of NW includes CO overtone absorption, Na and Ca absorption features, and an H$_2$ emission line, likely contamination from the outflow. In contrast, the spectrum of SE exhibits a stronger CO bandhead absorption, weaker Na and Ca absorption, and no emission lines. Br-$\gamma$ emission is not observed in either source, in contrast to near-IR spectra of many accreting young stars \citep{2010Connelley,2018Connelley}.

To constrain the NIR spectral type of SSTgbsJ214706, we use a $\chi^2$ minimization procedure to find the best match between the GNIRS spectrum and the medium-resolution stellar spectra provided in the IRTF spectral library \citep{2009Rayner} within the wavelength range spanning from 2.05 to 2.40 $\mu$m. The spectral characteristics of SE do not align perfectly with any specific spectral type. However, at the longer wavelength end, it resembles an M-type giant, primarily due to the enhanced CO absorption features, which are typical of FU Ori-type objects \citep{1996Hartmann}. The spectrum of the NW component closely resembles that of a G-type giant with a typical photosphere, exhibiting weaker CO absorption features when compared to K-M type giants and bonafide FU-Ori type objects (see Figure \ref{sp}). 

The SSTgbsJ214706 SE spectrum shows a strong resemblance to the standard spectrum of the bonafide FU Ori-type object, V2775\,Ori (Figure \ref{sp},\ref{co}). The CO absorption feature (at $\sim 2.29\,\mu$m) of the two are almost identical in terms of both depth and slope. Another weaker absorption feature is also observed at $\sim 2.35\,\mu$m, similar to V2775\,Ori, which is very faint for the NW component. Interestingly, this particular feature is commonly seen in the spectra of bondafide FUors \citep[see figure 5 in][]{2018Connelley}. However, what causes this feature is not clear. 

The equivalent widths of CO (measured from 2.292 to 2.320 $\mu$m), Na (2.208 $\mu$m), and Ca (2.265 $\mu$m) absorption features are calculated for both SSTgbsJ214706 NW and SE after convolving the GNIRS spectra (R=1700) with a Gaussian kernel to match the specified resolving power of R=1200 in \citet{2018Connelley}. The baseline is defined as a tangent connecting the mean local continuum on either side of the absorption feature and the equivalent widths are quantified by integrating the area between the spectrum and the baseline tangent.

 To ensure robustness, we employ a Monte Carlo approach involving 1000 iterations, introducing Gaussian noise to the flux spectrum with a standard deviation corresponding to the signal-to-noise ratio (SNR)\footnote{The average signal-to-noise ratio (SNR) for the full spectrum is derived using the DER\_SNR algorithm by \citet{2008SNR} using a wavelength range of 2.1 to 2.4 $\mu m $.}. We derive the mean and standard deviation of the resulting 1000 equivalent widths, which is adopted as the measured equivalent width and its associated 1$\sigma$ uncertainty. The equivalent widths obtained from extinction-corrected and non-extinction-corrected spectra of NW and SE are given in Table  \ref{table:2}.

\begin{figure}
    \centering
    \includegraphics[scale=0.35]{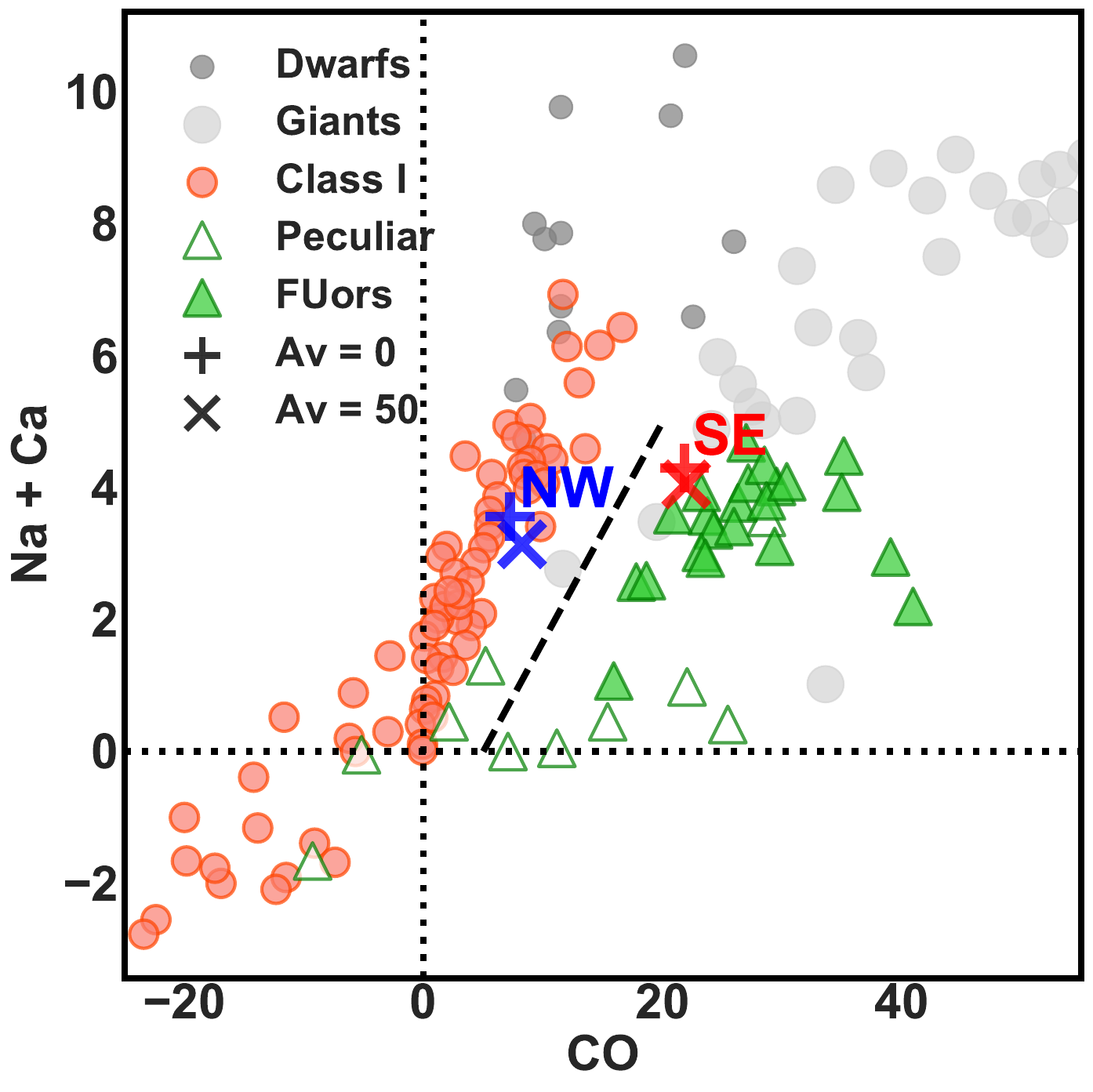}
    \caption{Equivalent width of CO ($2.29$\, $\mu$m) versus Equivalent width of Na ($2.208$\, $\mu$m) + Ca ($2.265$\, $\mu$m) for class I YSOs, FU Ori-type objects, Giants and Dwarfs reproduced using values from \citet{2018Connelley}. The EWs of CO, Na, and Ca for NW and SE are superimposed. The (+) symbols represent EWs measured without applying any extinction correction to the spectrum, while the (X) marks represent the EWs measured from extinction-corrected spectra using $A_V$ = 50 mag. SE conforms to the criteria signifying FU Ori-type objects, while the NW aligns with the characteristics expected of a typical young stellar object. }
    \label{cr}
\end{figure}

We reproduced the diagram presented in Figure 7 of \citet{2018Connelley}, with the CO, Na and Ca equivalent widths for NW and SE superimposed. As a validation step, we measured equivalent widths of these absorption features for the bonafide FU Ori-type object V2775\,Ori, which closely align with the measurements of \citet{2018Connelley}. The position of the SE component in the plot is consistent with bonafide FU Ori-type and FUor-like objects, whereas the location of NW corresponds to the region where class I YSOs are found (see Figure \ref{cr}). 

Accurate extinction measurements for this source are constrained by the limited availability of photometric data. However, the derived equivalent widths of both NW and SE components remain consistent regardless of whether we measure them without any extinction correction or if we apply an extinction correction using an approximate value of $A_V$ = 50 mag (see Appendix).  Both NW and SE retain their relative positions on the EW(CO) versus EW(Na + Ca) plot, suggesting that the equivalent widths are robust to uncertainty in extinction (Figure \ref{cr}).

\begin{table}
\centering
\begin{tabular}{lcccc}

\hline

Source & EW(CO)  & EW(Na) &  EW(Ca) \\
\hline

NW ($A_v = 0 $) & 7.23 $\pm$ 1.86 & 2.23 $\pm$ 0.88  & 1.33 $\pm$ 0.90\\
NW ($A_v = 50 $) & 9.24 $\pm$ 1.92 & 1.95 $\pm$ 0.93 & 1.51 $\pm$ 0.92\\
SE ($A_v = 0 $ )  & 21.82 $\pm$ 2.10  & 2.23 $\pm$ 1.06  & 2.01 $\pm$ 0.89\\
SE ($A_v = 50 $ ) & 21.97 $\pm$ 2.17 & 2.05 $\pm$ 1.04 & 2.00 $\pm$ 0.81\\
V2775Ori ($A_v = 0 $ ) & 25.29 $\pm$ 0.61 & 1.08 $\pm$ 0.39 & 1.61 $\pm$ 0.27 \\
V2775Ori ($A_v = 27.5 $ ) & 26.04 $\pm$ 0.69  & 1.61 $\pm$ 0.43 & 1.68 $\pm$ 0.32 \\

\hline
\end{tabular}
\caption{EWs of CO, Na, and Ca absorption features for NW and SE with and without extinction correction. The EW of V2775Ori is also included for comparison \citep{2018Connelley}}. 
\label{table:2}
\end{table}

\section{A low-luminosity FU Ori-type object?}

SSTgbsJ214706 appears to be one of the lowest luminosity YSOs which shows an FU Ori-type outburst. Our observations using Gemini/GNIRS and IRTF-Spex resolve SSTgbsJ214706 into two sources, presumably a binary system. Pre-outburst photometry and SED fitting indicate that the outbursting component SE of SSTgbsJ214706 is likely a low luminosity YSO with a pre-outburst luminosity of $\sim 0.23\,$L$_\odot$. 

Assuming an age of 0.5\,Myr and a total luminosity primarily from the photosphere with minimal contribution from accretion, the pre-outburst luminosity of $\sim 0.23\,$L$_\odot$ indicates an upper mass limit of $0.17\,$M$_\odot$ based on the \citet{2015Baraffe} models. Alternatively, considering the pre-outburst UKIDSS K band detection limit of 18.1 and assuming an extinction of 50 magnitudes (AV), the Baraffe models suggest a mass of 0.11\,M$_\odot$. Given that the source is most likely younger than 0.5\,Myr and that the luminosity is affected by accretion, the actual mass of the source could be considerably lower according to the evolutionary tracks.

During the long term outburst ongoing from 2010 till 2022, the SE component increased in WISE/W2-band flux by a factor of over 8, which results in an outburst luminosity of $\sim 0.95\,$L$_\odot$ \citep[following the correlation measured by][]{2020Contreras}. The high-amplitude mid-IR variability and the strong $^{12}$CO absorption in the spectrum of SSTgbsJ214706 SE is consistent with a classification of an FU Ori-type outburst. The $\sim$ 2 magnitude rise in brightness and the duration of the outburst is also consistent with the expectations from large changes in the accretion rate \citep{2013Scholz, 2022Hillenbrandb}.

\begin{figure}
    \centering  
    \includegraphics[scale=0.35]{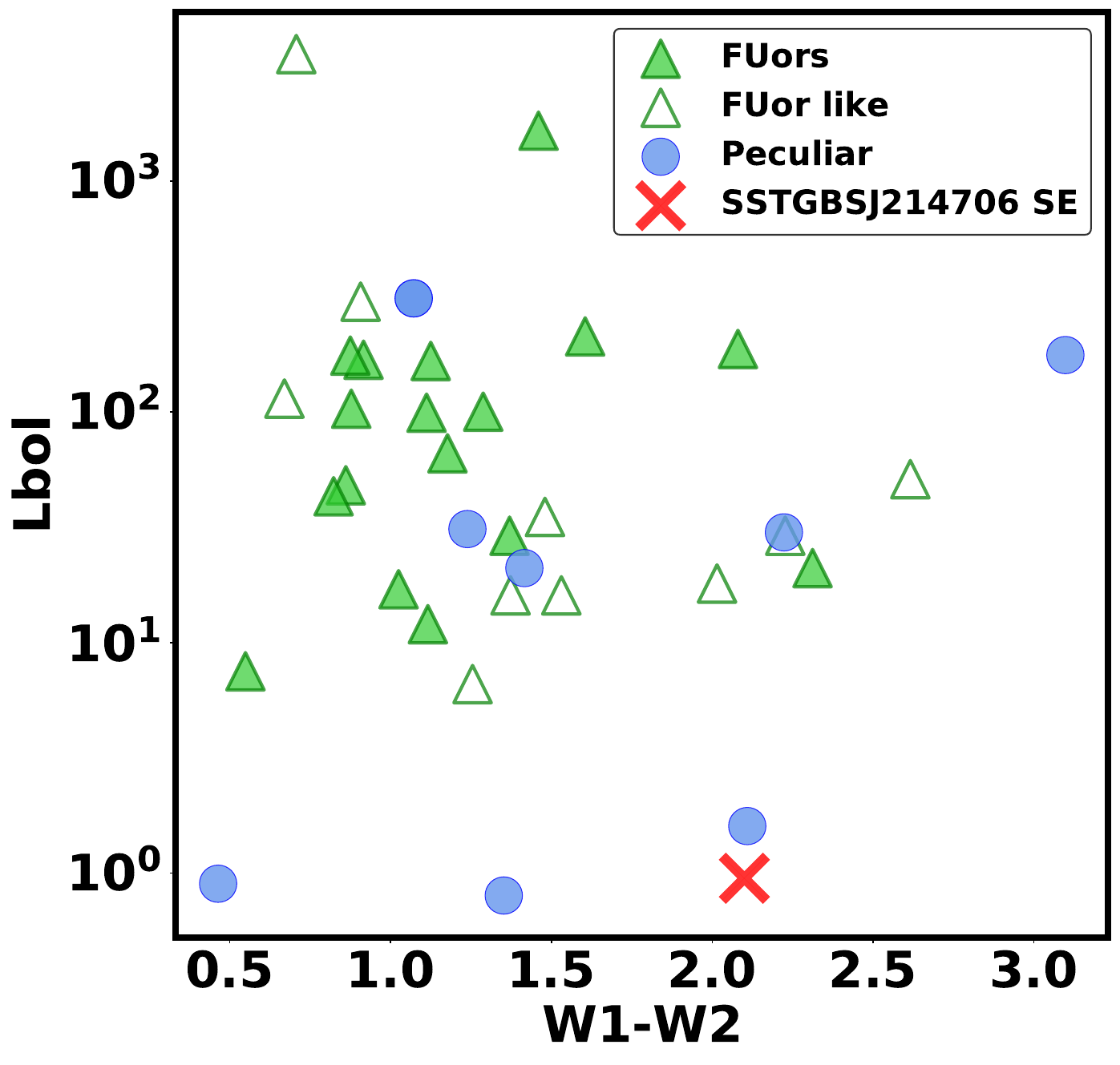}
    \caption{Luminosity distribution of bonafide FU Ori-type and FU Ori like objects. SSTgbsJ214706 SE shows a significantly lower outburst luminosity compared to previously observed FUors. }
    \label{lum}
\end{figure}

The outburst in SSTgbsJ214706 SE is of much lower luminosity than measured for previous and well-studied FU Ori-type outbursts (see Figure \ref{lum}). The 33 FUor and FUor-like sources listed in \citet{2018Connelley} show a median luminosity of 48\,L$_\odot$. In \citet{2018Connelley}, only three peculiar sources, IRAS 06393+0913, IRAS 18341-0113S and EC 53, have bolometric luminosities lower than 5\,L$_{\odot}$. \citet{2018Connelley} argue that some objects with low luminosities could actually be mid-to-late M-type objects, since both FUors and M-type objects have near-IR emission produced by a cool, optically thick surface. 

However, the concept of an FUor does not necessarily exclude low luminosity. The emergence of spectral features of the viscous disk only requires the latter to outshine the host star. Specifically, we may estimate the disk accretion luminosity as $L_{\rm acc}=GM_*\dot{M}/2R_*$ \citep{1973Shakura}, where $R_*$ is the stellar radius, and FUors require the ratio $\eta=L_{\rm acc}/L_*$ to be a factor of several or larger \citep[see also][where $\eta$ was defined in the near-IR]{2022Liu}. For young stars of age 0.5\,Myr, standard pre-main sequence evolution models give $L_*/L_\odot\sim3.3(M_*/$M$_\odot)^{1.5}$ and $R_*/R_\odot\sim3.5(M_*/$M$_\odot)^{0.5}$ \citep{2015Baraffe}, and thus
\begin{equation}
    \eta = \frac{GM_*\dot{M}}{2R_*L_*}\sim 1.4\left(\frac{M_*}{\rm 0.1\,M_\odot}\right)^{-1.0}\left(\frac{\dot{M}}{\rm 10^{-7}\,M_\odot\,yr^{-1}}\right)\ .
\end{equation}
For a star of $M_*=0.1\,$M$_\odot$ and $\dot{M}=7\times10^{-7}\,\rm M_\odot\,yr^{-1}$ we obtain $\eta=9.8$, so the viscous disk outshines the photosphere and should have an FUor spectrum.
Using $\eta=9.8$ results in $L_{\rm acc}=1.02 \,$L$_\odot$, comparable to the outburst luminosity of SSTgbsJ214706 SE. A much lower accretion luminosity could also be considered in the case of a viscously heated disk around a protoplanet \citep{2015Zhu}. 

\subsection{An alternative interpretation: a reduction in extinction}

A possible alternative interpretation for SSTgbsJ214706 SE would be that it is a young M-type object (very low-mass star or brown dwarf) or a background M-giant undergoing a change in extinction. Throughout most (but not all) of the brightness increase of SSTgbsJ214706 SE, the colour changes follow the reddening vector, consistent with $\Delta A_{V}=63$~mag \citep[following the extinction law of ][]{2019Wang}.

High amplitude ($\Delta K_{s}>4$~mag) extinction-related events have been observed in giant stars \citep{2021Smith}, with light curves that can resemble long-term outbursts (Lucas et al.,~submitted). The spectra taken at their peak brightness show strong $^{12}$CO (beyond 2.29 $\mu$m) and $^{13}$CO (beyond 2.34 $\mu$m) first overtone absorption (\citealp{2021Guo}, Guo et al.~submitted). The EWs of $^{12}$CO of these variable giant stars \citep[e.g. VVVv319 in][with EW = 57 \AA]{2021Guo} puts them in the upper right region of Fig. \ref{cr}, beyond the location of FUors and of SSTgbsJ214706 SE. Our source does show an absorption feature at $\sim$2.345 $\mu$m (Fig. \ref{co}) which is close to the wavelength of the first bandhead of $^{13}$CO. However, no additional absorption from $^{13}$CO seems apparent in the spectrum of our source. Therefore, we conclude that the absorption feature at $\sim$2.345 $\mu$m is more likely related to a similar feature observed in FUors rather than to $^{13}$CO in the spectrum of a giant star.

In YSOs, obscuration by structural inhomogeneities in the circumstellar disk lead to variability with timescales that depend on the location of the structure in the disk, and with amplitudes that are effectively limitless and wavelength-dependent \citep{2001Carpenter,2013Hillenbrand}. If extinction drives the variability of SSTgbsJ214706 SE, then the change of $\Delta A_{V}=63$~mag  is much larger than known extinction-related changes of AA Tau \citep[$\Delta A_{V}=2$~mag and $\Delta(W1)\sim1$~mag,][]{2013Bouvier,2021Covey} or V2492 Cyg \citep[$\Delta A_{V}=35$~mag,][]{2013Hillenbrand,2023Carlos}. Periodic ($P<200$ d) achromatic changes due to obscuration by a circumbinary disk can reach large near-IR amplitudes \citep[$\Delta K\sim3$ mag,][]{2020Garcia,2022Zhu}. High amplitudes ($\Delta K_{s}>4$~mag) have been also observed in a sample of extinction-related variable YSOs from the VVV survey (Lucas et al., submitted). However these events often appear in the light curve as long-term dips that last no longer than 1500 d. The light curve of SSTgbsJ214706 SE does not show any periodicity or long-term dips that could be related to this type of obscuration events.

Some of the mid-IR brightening must be unrelated to extinction, since the colour change of SSTgbsJ214706 does not follow the reddening vector at all times in the light curve.  The colour-magnitude diagram (Figure \ref{lc}) shows an ``s-shape'' curve, with the brightening at the end of the burst corresponding with redder emission.  At the start of the burst, the brightening also corresponds to redder colors, likely because of a change in the relative contributions between the NE and SW components.

\subsection{The slow rise of SSTgbsJ214706 SE}

Light curves of FU ori-type objects display diverse rise and peak durations, ranging from months to years; and many have rise times that are poorly constrained \citep{2022Fischer}. The e-folding time ($\sim$ 7 years) measured for the increase in brightness of SSTgbsJ214706 SE is significantly longer than the rise time (at optical wavelengths) of the majority of other known FU Ori outbursts, which typically exhibit shorter e-folding time scales for photometric brightening of less than one year (e.g. FU Ori \citealt{2000Kenyon}; V1057 Cyg \citealt{2021Szab}; V2775 Ori \citealt{2012Fischer}; V960 Mon \citealt{2023Carvalho}; HBC 722 \citealt{2010Semkov}). The prototype, FU Orionis, exhibited an e-folding time of approximately 50 days during its rising phase \citep{1977Herbig, 2000Kenyon}, 50 times shorter than the rise of SSTgbsJ214706. 

However, outbursts such as V1515 Cyg \citep{2022Szabo}, V1735 Cyg \citep{2020Wendeborn}, V733 Cep \citep{2010Peneva}, and V900 Mon \citep{2021Semkov} show that the rise time can extend up to 20--30 years. The long timescale accretion variability is expected to be driven by the mismatch between steady-state mass flow through the disk and local regions where the flow becomes stalled \citep{2022Fischer}. We propose that the enhancement in brightness observed in SSTgbsJ214706 is influenced by a large increase in the accretion rate, characterized by a prolonged e-folding time, attributed to its specific location within the disk where the instability initiates. This hypothesis aligns with the long timescales of luminosity enhancement in near- and mid-infrared, as illustrated in the models of FU Ori-type objects in \citet{2023Cleaver}. Further observations and research into SSTgbsJ214706 could yield valuable insights into the mechanisms driving episodic accretion in low-mass embedded YSOs.

\section{Summary \& Conclusions}

SSTgbs J21470601+4739394, a low luminosity YSO associated with IC 5146, has displayed a gradual and continuing increase in mid-IR brightness since 2010, as observed with NEOWISE. To characterize the nature of this outburst, we conducted a photometric analysis, subsequently complemented by Gemini near-infrared spectroscopy. Our investigations reveal that this object comprises of two distinct components and its south-east (SE) component is presently undergoing an outburst similar to that of the FU Ori phenomenon. The key outcomes of our study are as follows:

\begin{enumerate}
    \item The mid-IR emission from SSTgbsJ214706 shows a gradual brightening of 1.78 mag at 3.6 (W1) and 1.84 mag at 4.5 $\mu$m (W2) over the past 12 years, with an e-folding rise time of $\sim 7$ years. 
    \item The aquisition images taken during follow-up spectroscopy with Gemini and IRTF resolve SSTgbsJ214706 into two point sources, the NW and SE components, along with a spatially extended outflow. Comparison with pre-outburst UKIDSS images and the shift in the centre of brightness of NEOWISE W2 images indicate that the outburst is occurring on the more embedded SE component, which dominates the mid- and far-infrared emission from the source. 
    \item The outbursting component SE has a $K$ band spectrum consistent with FU Ori-type outbursts. The spectral characteristics, particularly the existence of enhanced absorption in CO bands from $2.29\,\mu$m, closely resemble the spectra of bonafide FUors V2775 Ori and FU Ori. The $K$ band spectrum of the NW component matches that of a G star and does not appear to be strongly accreting. 
    \item  The luminosity of the SE component is estimated to be 0.23\, L$_\odot$ before the outburst and 0.95\,L$_\odot$ during the outburst, which is 1–2 orders of magnitude fainter than bonafide FU Ori outbursts. 
    \item From the color-magnitude diagram and spectral features, we conclude that the main mechanism driving the variability of SSTgbsJ214706 is changes in the accretion rate of the system, although some contribution from extinction variability cannot be discarded.
\end{enumerate}

\section*{Acknowledgement}
We thank the referee for providing helpful comments.
JJ acknowledges the financial support received through the DST-SERB grant SPG/2021/003850.
CCP was supported by the National Research Foundation of Korea (NRF) grant funded by the Korean government (MEST) (No. 2019R1A6A1A10073437). DJ is supported by NRC Canada and by an NSERC Discovery Grant.
JEL was supported by the New Faculty Startup Fund from Seoul National University and the NRF grant funded by the Korean government (MSIT) (grant number 2021R1A2C1011718).

This research has made use of the NASA/IPAC Infrared Science Archive, which is funded by the National Aeronautics and Space Administration and operated by the California Institute of Technology. This work is based in part on data obtained as part of the UKIRT Infrared Deep Sky Survey. 

This research has made use of the VizieR catalogue access tool, CDS, Strasbourg, France (DOI : 10.26093/cds/vizier). The original description of the VizieR service was published in 2000, A\&AS 143, 23.

\section*{Data Availability}

The photometric data underlying this article are available in the Appendix. The spectra will be shared on reasonable request to the corresponding author.

\bibliographystyle{mnras}
\bibliography{ref_1.bib} 




\section{Appendix}
\subsection{Extinction estimates}

We carried out evaluations to estimate the extinctions of NW and SE components of SSTgbsJ214706 separately. For the NW component, we make the assumption that it is a non-variable source and estimate an extinction of approximately $A_V \sim 50$ mag using the following method. Assuming a spectral type of G7III, the intrinsic K - W1 value was obtained as 0.09 from \citet{2013Pecaut}. The IRAC1 and IRAC2 values for NW were derived through PSF photometry conducted on Spitzer images, as described in section \ref{sec:photo}. These values were then approximated as W1 and W2. The K band magnitude is derived from the UKIDSS DR6. The excess in observed K-W1 was converted into $A_V$ using the relations from \citet{2019Wang}.

\begin{figure}
    \centering
    \includegraphics[scale=0.35]{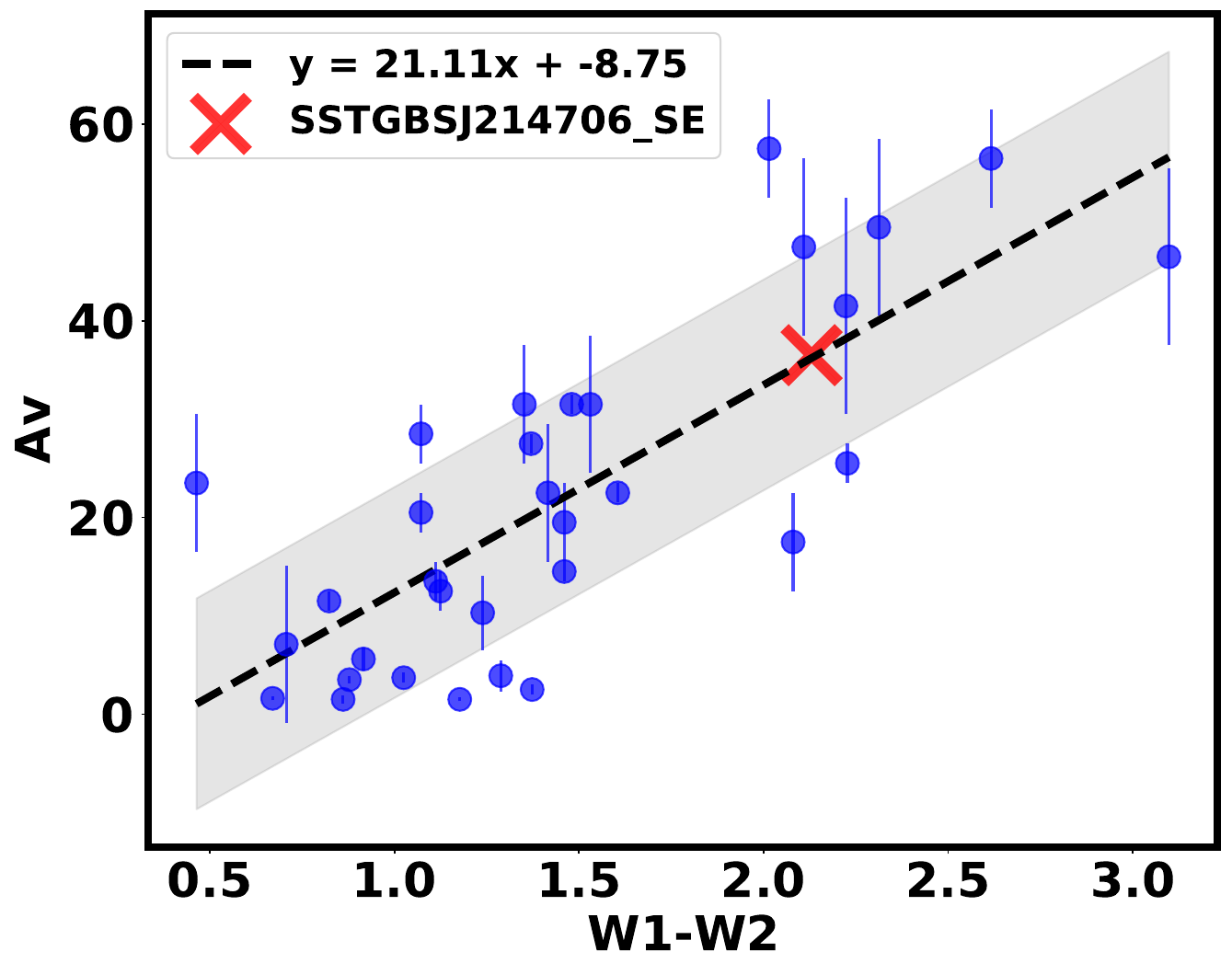}
    \caption{The relationship between the W1-W2 color and $A_V$ for bonafide FUors and FUor-like objects. W1 and W2 magnitudes are sourced from \citet{2012Cutri}, while $A_V$ values are obtained from \citet{2018Connelley}.}
    \label{correlation}
\end{figure}

For SE, due to its variable nature, the aforementioned approach is less suitable. Consequently, we established an empirical correlation between the observed W1-W2 colors and estimated $A_V$ values for bonafide FUors and FUor-like objects from \citet{2018Connelley} (see Figure \ref{correlation}). Following this correlation, the W1-W2 color of SE corresponds to an $A_V$ range of $\sim 25 - 50$. It is important to note that both of these methods are subject to large uncertainties stemming from the various assumptions and approximations employed. 

\begin{figure}
   \centering
    \includegraphics[scale=0.35]{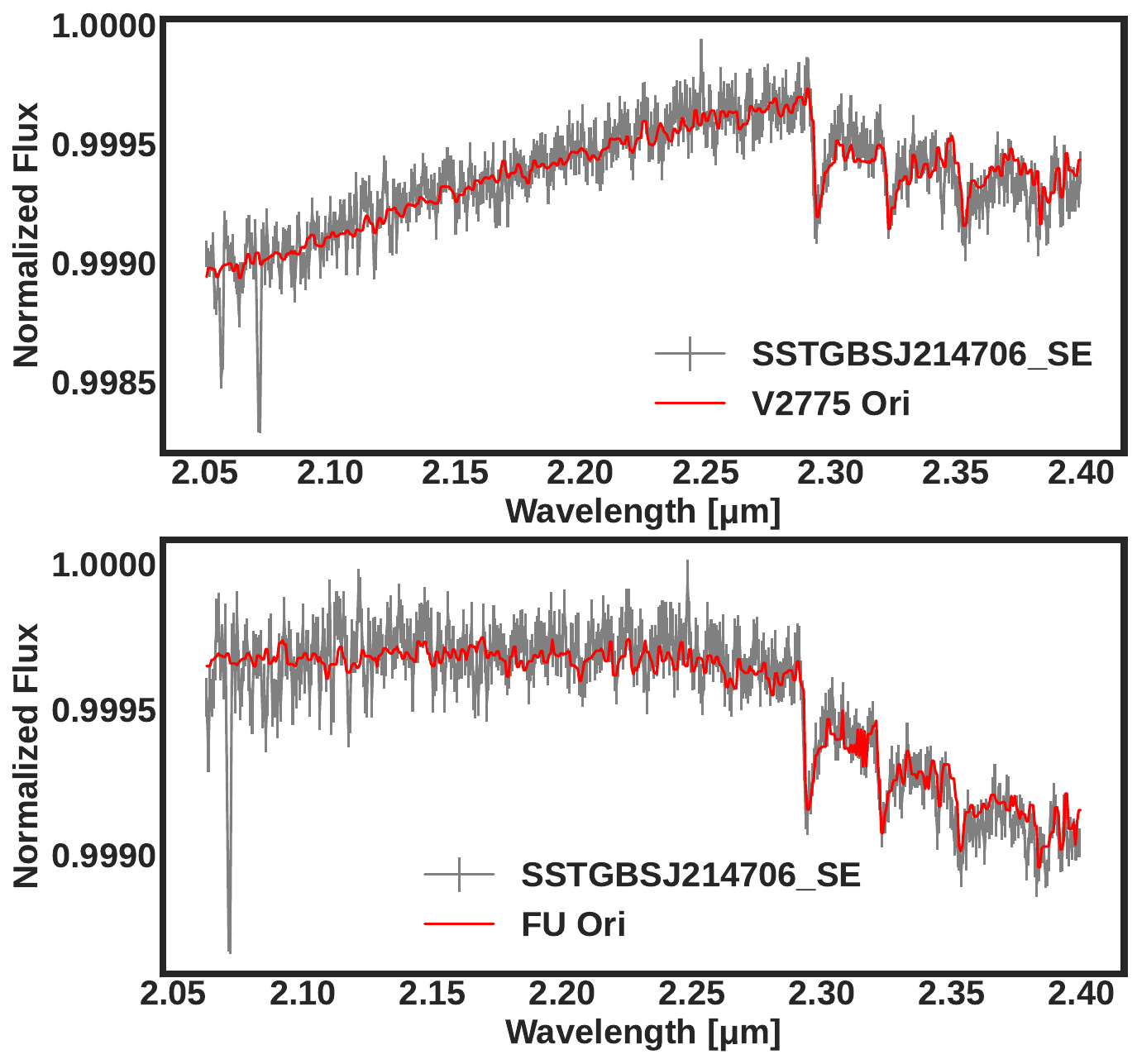}
    \caption{ SSTgbsJ214706 SE spectra fitted with FU Ori and V2775 Ori templates with variable extinction parameter ($A_v$), revealing a strong resemblence to bonafide FU Ori-type objects.}
    \label{fuori}
\end{figure}

We also employed the methodology outlined in \citet{2018Connelley}, attempting a continuum fit of the spectrum with the spectra of FU Orionis and V2775 Ori for SE (see Figure \ref{fuori}). This procedure yielded an approximate value of $\sim 58$ magnitudes for the extinction. It is important to acknowledge that this approach is also affected by inaccuracies stemming from the flux calibration uncertainties present in our GNIRS spectrum.

 While these methods yield $A_V\sim50$ mag for both components, the SEDs demonstrate that the SE component must be much more embedded and have a larger extinction than the NW component.

\begin{figure}
    \centering
    \includegraphics[scale=0.35]{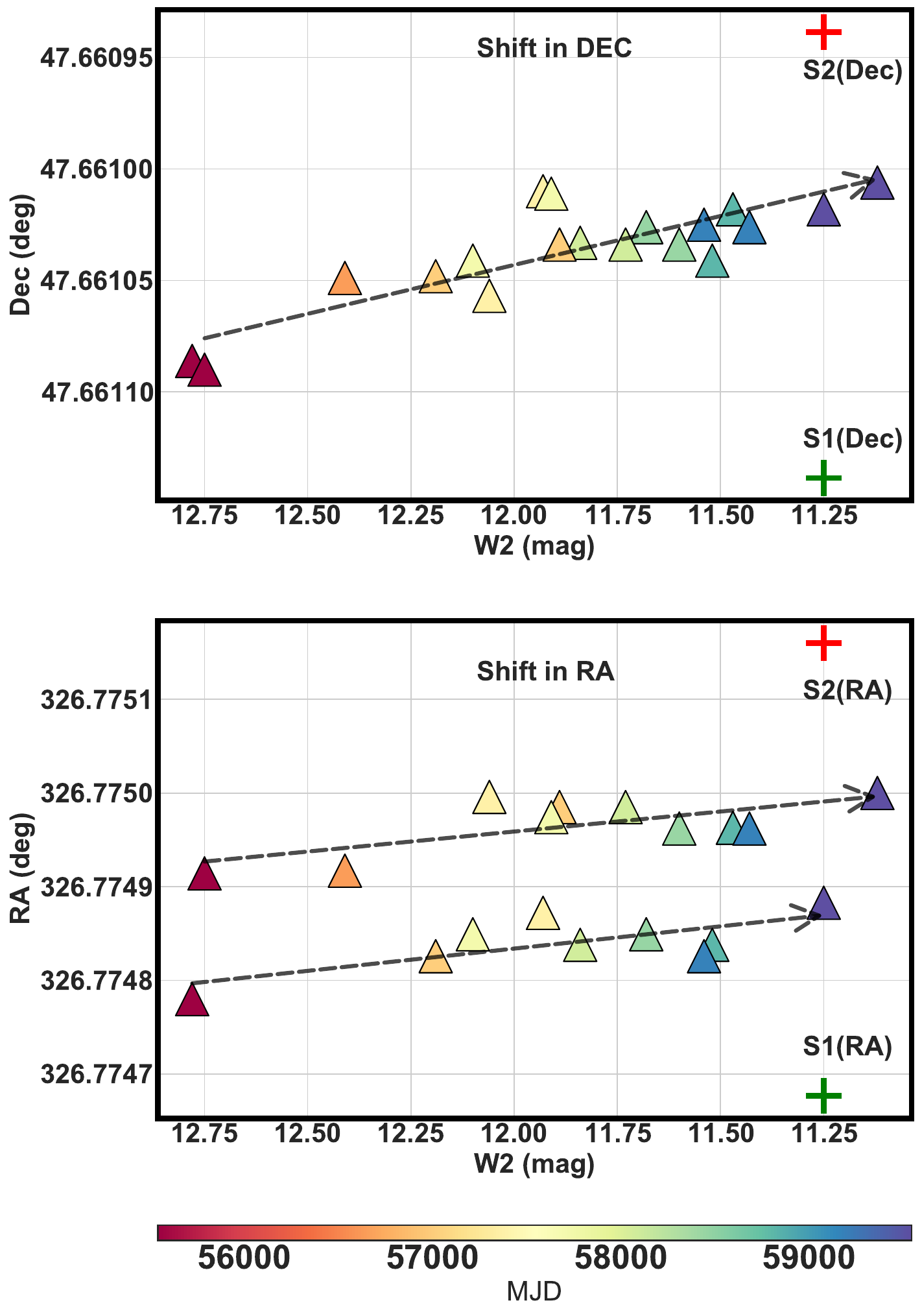}
    \caption{ Systematic shift of center of brightness of the system towards the location of SE in ra and dec as observed in NEOWISE W2 images from 2010 to 2022.}
    \label{sys}
\end{figure}

\begin{table*}
\centering

\begin{tabular}{lcccc}

\hline

Epoch (MJD) & Band & NW Flux (mJy) & SE Flux (mJy)& Combined Flux (mJy)\\
\hline

53669 & UKIDSS K & 0.21 & - & - \\
59782 & IRTF/Spex K &  0.27  & 0.23 & - \\
54065 & Spitzer IRAC 3.6 & 0.608 & 0.332 & 0.700 \\
54065 & Spitzer IRAC 4.5 & 0.974 & 1.200 & 1.500\\
54065 & Spitzer IRAC 5.8 & 0.459 & 1.800 & 1.900\\
54065 & Spitzer IRAC 8.0 & - & 1.700& - \\
53319 & Spitzer MIPS 24 &- & 18.00 & - \\

55366 & AllWISE W1 &- &- & 0.405  \\
55546 & AllWISE W1 & -& -& 0.429  \\
56641 & NEOWISE W1 &- &- & 0.533 \\
56826 & NEOWISE W1 & -&- & 0.658 \\
57006 & NEOWISE W1 &- & -& 0.960 \\
57190 & NEOWISE W1 &- &- & 0.891 \\
57365 & NEOWISE W1 &- & -& 0.840 \\
57555 & NEOWISE W1 &- & -& 0.808 \\
57724 & NEOWISE W1 & -&- & 0.892 \\
57919 & NEOWISE W1 &- & -& 1.009 \\
58087 & NEOWISE W1 & -& -& 1.192 \\
58286 & NEOWISE W1 & -&- & 1.346 \\
58448 & NEOWISE W1 & -&- & 1.405 \\
58650 & NEOWISE W1 & -&- & 1.568 \\
58812 & NEOWISE W1 &- & -& 1.620 \\
59017 & NEOWISE W1 & -&- & 1.705 \\
59176 & NEOWISE W1 & -& -& 1.716 \\
59381 & NEOWISE W1 & -& -& 2.017 \\
59544 & NEOWISE W1 & -& -& 2.035 \\
59745 & NEOWISE W1 & -& -& 2.013 \\
59908 & NEOWISE W1 & -& -& 2.201 \\

55366 & AllWISE W2 &- &- & 1.320 \\
55546 & AllWISE W2 & -&- & 1.363 \\
56641 & NEOWISE W2 & -& -& 1.850 \\
56826 & NEOWISE W2 &- &- & 2.277 \\
57006 & NEOWISE W2 & -& -& 3.007 \\
57190 & NEOWISE W2 &- & -& 2.887 \\
57365 & NEOWISE W2 & -&- & 2.559 \\
57555 & NEOWISE W2 & -&- & 2.476 \\
57724 & NEOWISE W2 &- &- & 2.941 \\
57919 & NEOWISE W2 &- &- & 3.114 \\
58087 & NEOWISE W2 & -& -& 3.491 \\
58286 & NEOWISE W2 & -& -& 3.639 \\
58448 & NEOWISE W2 & -&- & 3.931 \\
58650 & NEOWISE W2 & -&- & 4.201 \\
58812 & NEOWISE W2 & -&- & 4.411 \\
59017 & NEOWISE W2 & -&- & 4.117 \\
59176 & NEOWISE W2 & -& -& 4.460 \\
59381 & NEOWISE W2 & -& -& 5.352 \\
59544 & NEOWISE W2 & -& -& 6.089 \\
59745 & NEOWISE W2 & -& -& 6.208 \\
59908 & NEOWISE W2 & -& -& 6.763 \\

55366 & AllWISE W3 & -& 0.445 &-  \\
55546 & AllWISE W3 &- & 0.322 & -\\
55366 & AllWISE W4 &- & 12.905 & - \\
55546 & AllWISE W4 & -& 15.832& -\\
55345 & Herschel PACS 70 &- & 134.6 & - \\
55909 & Herschel PACS 100 & -& 262.2 &-  \\
56177 & JCMT SCUBA 450 & -& 340.0& - \\
56177 & JCMT SCUBA 850 & -& 74.0&- \\

\hline

\end{tabular}
\caption{Compilation of all the photometric data from various surveys for the object SSTgbsJ214706. WISE W1, W2 are the combined fluxes for the unresolved system \citep{2014Cutri}. UKIDSS DR6 K band photometry are from \citep[GPS,][]{2008Lucas}. Spitzer/IRAC photometry for the combined system are obtained from \citep{2008Harvey}. Individual fluxes at 3.6, 4.5 and 5.8 $\mu$m are estimated using PSF photometry of the extended source in spitzer IRAC images. Herschel PACS photometry data were sourced from \citep{2016Kim}, and JCMT SCUBA-2 450 and 850 $\mu$m data were obtained from \citep{2017Johnstone}. All fluxes above $8 \mu$m are assumed to be arising from  SE alone. The WISE and UKIDSS magnitudes are converted into fluxes by applying the zeropoints: W1;309.54, W2;171.79, W3;31.67, W4;8.36 Jy \citep{2010Wright}, UKIDSS:K;631 Jy \citep{2006Hewett}.}


\label{table:1}

\end{table*}


\begin{table}
	\centering
	\caption{Source parameters obtained from SED fitting.}
	\label{sedparams}
\begin{tabular}{|c|c|} 
    \hline
        Model name & 6gil2x06\_01\\
        Av & 65\\
        Star radius (R$_{\odot}$) & 3.92\\
        Star temperature (K) & 4103\\
        Disk mass (M$_{\odot}$) &2.021$e-02$ \\
    \hline
\end{tabular}
\end{table}

\bsp	
\label{lastpage}
\end{document}